\documentclass[conference]{IEEEtran}
\IEEEoverridecommandlockouts
\usepackage{cite}
\usepackage{amssymb}
\usepackage{amsmath,amsfonts,amsthm,bm}
\usepackage{mathtools}
\usepackage{algorithmicx}
\usepackage{graphicx}
\usepackage{textcomp}
\usepackage{xcolor}
\usepackage{algorithm,caption}
\usepackage{comment}
\usepackage{dsfont}
\usepackage{subcaption}

\usepackage{algorithm}
\graphicspath{{images/}}

\newcommand{\RNum}[1]{\uppercase\expandafter{\romannumeral #1\relax}}
\def\BibTeX{{\rm B\kern-.05em{\sc i\kern-.025em b}\kern-.08em
    T\kern-.1667em\lower.7ex\hbox{E}\kern-.125emX}}

\newtheorem{definition}{Definition}

\begin{document}

\title{Deep Reinforcement Learning for Random Access in Machine-Type Communication\\

}

\author{\IEEEauthorblockN{Muhammad Awais Jadoon\IEEEauthorrefmark{1},
Adriano Pastore\IEEEauthorrefmark{1},
Monica Navarro\IEEEauthorrefmark{1}, and 
Fernando Perez-Cruz\IEEEauthorrefmark{2}} 
\IEEEauthorblockA{\IEEEauthorrefmark{1}Centre Tecnològic Telecomunicacions Catalunya (CTTC)/CERCA, Castelldefels, Spain}
\IEEEauthorblockA{\IEEEauthorrefmark{2}Swiss Data Science Center (ETH Zurich and EPFL) and Computer Science Department ETH Zurich, Switzerland }\
Email: \{mjadoon, apastore, mnavarro\}@cttc.es, fernando.perezcruz@sdsc.ethz.ch 

}

\bstctlcite{IEEEexample:BSTcontrol}

\maketitle

\begin{abstract}
Random access (RA) schemes are a topic of high interest in machine-type communication (MTC). In RA protocols, backoff techniques such as exponential backoff (EB) are used to stabilize the system to avoid low throughput and excessive delays. However, these backoff techniques show varying performance for different underlying assumptions and analytical models. Therefore, finding a better transmission policy for slotted ALOHA RA is still a challenge. In this paper, we show the potential of deep reinforcement learning (DRL) for RA. We learn a transmission policy that balances between throughput and fairness. The proposed algorithm learns transmission probabilities using previous action and binary feedback signal, and it is adaptive to different traffic arrival rates. Moreover, we propose average age of packet (AoP) as a metric to measure fairness among users. Our results show that the proposed policy outperforms the baseline EB transmission schemes in terms of throughput and fairness.
\end{abstract}


\section{Introduction}
Random access (RA) protocols such as slotted ALOHA are highly relevant in designing the multiple access schemes for massive machine-type (MTC) in future wireless networks. In slotted ALOHA RA protocol \cite{slotted-aloha}, retransmission strategies are usually employed to resolve collisions and to optimize metrics such as throughput, fairness or delay.
There exist several widely used strategies that exploit feedback signals from the receiver to implement a transmission control mechanism \cite{Hajek82, kumar1984, massey_RA}. One of the widely used transmission strategies is exponential backoff (EB). In EB, the probability of (re)transmission is multiplicatively decreased by a backoff factor $\sigma$ each time a collision occurs. Optimal values for the backoff factor $\sigma$ and different flavours of the backoff policy have been addressed in the literature.  Binary Exponential Backoff (BEB), i.e., when $\sigma=2$, has been used in standards such as Ethernet LAN, IEEE 802.3 or IEEE 802.11 WLAN and it has been considered in \cite{beb_analysis, goodman_1988} for theoretical analyses. Recently, it has been shown that $\sigma = 1.35$ performs better in terms of delay compared to BEB \cite{Barletta18}. An algorithm similar to BEB has been used in \cite{Hajek82} that considers past feedback to adjust transmission probability, but such algorithm requires constant sensing of the channel. 

We divide EB schemes operating with packet collision feedback into non-symmetric versus symmetric exponential backoff (nSEB and SEB), respectively. In nSEB only the users that suffer the collision back off, whereas in the SEB, all users back off when a collision occurs, regardless of whether they attempted transmission or not. Different assumptions such as packet arrival traffic model, feedback type (e.g. binary or ternary feedback signaling) influence the performance of these schemes. For instance, nSEB leads to the well-known \textit{capture effect}, where a single user or a reduced set of users occupies all channel resources. Analytical closed-form solutions for such well-studied schemes are still an open research problem depending on system model assumptions, such as  stability and queue assumptions.
When considering multiple access for the MTC in 5G and beyond communication technologies, these protocols cannot be directly applied and therefore, machine learning tools appear attractive to model multiple access problems.

In recent years, deep reinforcement learning (DRL)
has attracted much attention in wireless communication research as a potential tool to model multiple access problems. The main motivation for using RL is its ability to learn near-optimal strategies by interacting with the environment through trial-and-error. The application of RL for channel access goes back to 2010 where Q-learning was used for a multi-agent RL setting \cite{HLi}. In \cite{CHU201523}, ALOHA-Q protocol is proposed for a single channel slotted ALOHA scheme that uses an expert-based approach in RL. The goal in that work is for nodes to learn in which time slots the likelihood of packet collisions is reduced. However, the ALOHA-Q depends on the frame structure and each user keeps and updates a policy for each time slot in the frame. In \cite{alfaro_alohaq}, the ALOHA-Q is enhanced by removing the frame structure. However,  every user still has to keep the number of policies equal to the time slots window it is going to transmit in. Other works such as \cite{Naparstek, zhong2019deep, wang2018} consider RL-based multiple access works for multiple channels. In \cite{Naparstek} and \cite{wang2018} deep Q-Network (DQN) algorithm is used for multiple user and multiple channel wireless networks. As opposed to \cite{Naparstek}, we use a different and a smaller set of state parameters, i.e., we consider only one previous action and feedback. In \cite{zhong2019deep}, another DRL algorithm known as \textit{actor-critic} DRL is used for dynamic channel access. Furthermore, in \cite{8Yu2019het}, a heterogeneous environment is considered in which an RL agent learns an access scheme in co-existence with slotted ALOHA and a time division multiple access (TDMA) access schemes. 

In this work, we leverage on DRL to design a channel access and transmission strategy for RA, aiming for reduced signaling and no user coordination. More specifically, the system model considers a binary broadcast feedback, common to all users. We learn a policy that can be designed to be equal for all users through centralised training or it can be optimised or adapted individually through online training. In this particular work, we have considered the former approach. We consider Poisson process for packet arrivals; however, as opposed to the works mentioned above, our proposed scheme and results are not constrained to the specific arrival process. Furthermore, unlike all the above-mentioned works, we do not assume the system to be in \textit{saturation state} (when users always have a packet in their buffer). This state is of particular interest in MTC systems where not all users will always have a full buffer because of the sporadic traffic arrivals.

The rest of the paper is organised as follows: Section \ref{sec:system_model} introduces the system model and defines the performance metrics, Section \ref{sec:rl_env} describes the RL environment and  DQN architecture, simulation results are provided in Section \ref{sec:experiments} and finally conclusions are drawn in Section \ref{sec:conclusion}. 

\section{System Model and Problem Formulation}\label{sec:system_model}
We consider a synchronous slotted RA system with a set $\mathcal{N} = \{1, \dotso, N \}$ of active users, a central receiver such as a base station, and an error-free broadcast channel. The physical time is divided into slots, with slot index $k \in \mathbb{N}$, and the duration of each slot is normalized to $1$. We assume that every packet spans exactly one time slot, and all users are perfectly synchronized. Each user is equipped with a buffer and we assume that it can store one packet. The buffer state of user $n$ at time $k$ is defined as $B_n(k) \in \{0,1\}$, where $B_n(k)=1$ if there is a packet in the buffer and it is $0$ otherwise. If buffer $B_n(k)$ is full, new packets arriving at user $n$ are discarded and are considered lost. At each time slot $k$, user $n$ takes an action $A_n(k) \in \{0,1\}$, where $A_n(k)=0$ corresponds to the event when user $n$ chooses to not transmit and $A_n(k)=1$ corresponds to the event when user $n$ transmits a single packet on the channel. If only one user transmits on the channel in a given time slot, the transmission is successful, whereas a collision event happens if two or more users transmit in the same time slot. The collided packets are discarded and need to be retransmitted until they are successfully received. We define the binary feedback signal $F(k) \in \{0,1\}$, which is broadcast to all users, as
\begin{align}
    F(k) = 
    \begin{cases}
           0 & \text{if there is a collision in time slot $k$} \\
           1 & \text{otherwise.}
    \end{cases}
\end{align}
Moreover, we assume that each user keeps a record of its previous action $A_n(k-1)$, the previous feedback $F(k-1)$ from the receiver and its current buffer state $B_n(k)$. We refer to the tuple
\begin{align}
    S_n(k) = ( A_n(k-1), F(k-1), B_n(k)) \label{eq:history_user_n}
\end{align}
as the \emph{history} or \emph{state} of user $n$ at time $k$, and to $S(k) = (S_1(k),\dotsc,S_N(k))$ as the \emph{global history} of the system.
\subsection{Slotted Events}
Within time $k$, the events happen in the following order at user $n$:
\begin{enumerate}
    \item The buffer state of user $n$ is $B_n(k)$.
    \item A number $U_n(k)$ of new packets arrive. We assume that packet arrivals follow mutually independent Poisson processes with average arrival rate per user $\lambda_n$, where $\lambda_n = \lambda/N$ and $\lambda$ is the total arrival rate.
    \item   The buffer state $B_n(k)$ is updated to an intermediary buffer state as,
    $\tilde{B}_n(k) = \min\{ B_n(k) + U_n(k) , 1 \}$
    to account for the newly arrived packet.
    \item   If there is at least one packet in the buffer ($\tilde{B}_n(k) = 1$), the action $A_n(k)$ is drawn at random from the distribution $\pi_n(\cdot|S_n(k))$. Otherwise, $A_n(k) = 0$.
    \item   The feedback signal $F(k)$ is broadcast from the receiver to all $N$ users. 
    \item Based on the feedback signal observed by each user, if a packet has been transmitted successfully, i.e., $F(k)=1$ and $A_n(k)=1$), then the packet is deleted from the buffer. The buffer state $B_n(k+1)$ is updated as, $B_n(k+1) = \tilde{B}_n(k)- G_n(k)$,
    where $G_n(k) \in \{0,1 \}$ is the random variable that indicates when a packet from user $n$ has been successfully delivered to the receiver during the time slot $k$, and it is defined as,
    \begin{align} \label{eq:binary_feedback1}
        G_n(k) &= 
        \begin{cases}
    		1 & \text{if $A_n(k) = 1$ and $F(k)=1$}\\
    		0 & \text{otherwise.}
    	\end{cases} 
    \end{align}
        \item The values $A_n(k)$, $F(k)$ and $B_n(k+1)$ are used to update the history for the next time slot, $S_n(k+1)$.
    \end{enumerate}
\begin{definition}  
A policy or access scheme of user $n$ at time slot $k$, is a mapping from $S_n(k)$ to a conditional probability mass function $\pi_n(\cdot|S_n(k))$ over the action space $\{0, 1\}$. We consider a distributed setting in which there is no coordination or message exchange between users for the channel access. 
Each new action $A_n(k) \in \{0, 1\}$ is drawn at random from $\pi_n(\cdot|S_n(k))$ as follows:
\begin{equation}
    \mathrm{Pr}\bigl\{ A_n(k) = a \bigm| S_n(k) = s \bigr\}
    = \pi_n(a|s).
\end{equation}
\end{definition}
We are interested in developing a distributed transmission policy for slotted RA that can effectively adapt to changes in the traffic arrivals and provide better performance than the baseline reference EB schemes.
\subsection{Performance Metrics}
The objective is to evaluate a scheme that efficiently utilizes the channel resources as well as accounts for fairness between users. We consider throughput and propose a new metric, \textit{age of packet} (AoP), to measure fairness.\footnote{Jain's index~\cite{Jain1998AQM}, which is often used as a fairness metric in other publications~\cite{alfaro_alohaq}, is not adequate here: while our setting is perfectly fair on the long run (because policies and the channel model are perfectly symmetric among users), we are interested in quantifying the amount of capture effect as an indicator for short-term imbalances, which we interpret as a lack of fairness.}

\subsubsection{Throughput}
The channel throughput is defined as the average number of packets that are successfully transmitted from all users. For the finite time horizon $K$, and a given total arrival rate $\lambda$, the throughput is computed as
\begin{align}
    T = \frac{1}{K}\sum_{k=1}^K \sum_{n} G_n(k).
\end{align}
\subsubsection{Age of Packet (AoP)}
Certain policies like nSEB notoriously tend to be low in fairness, in that they suffer from the so-called \emph{capture effect}, by which one node occupies the channel for a large stretch of time, during which other nodes maintain a very low transmit probability and accumulate large delays. We quantify this effect via the average AoP.\footnote{This metric has a different connotation to age of information (AoI).} A low average AoP is thus indicative of fairness.
The AoP of user $n$, denoted as $w_n(k)$, grows linearly with time if a packet stays in the user buffer, and it is reset to $0$ if the packet is transmitted successfully. Specifically, we assume that $w_n(1)=0$, and the AoP $w_n(k)$ evolves over time as follows:
\begin{align}
    w_n(k) &=
    \begin{cases}
        0  &\text{if } B_n(k) =0 \\
        w_n(k-1) + 1 & \text{otherwise.}
    \end{cases}
\end{align}
The average AoP for user $n$ after a time span of $K$ time slots is given by
\begin{align}\label{eq:aop_n}
    \Delta_n = \frac{1}{K}\sum_{k=1}^{K} w_n(k)
\end{align}
and the average AoP of the overall system by $\Delta = \sum_n \Delta_n$.
For systems with a finite buffer size like ours, another relevant metric is the \textit{packet discard rate}. Due to space constraints, we do not focus on it in this paper, and do not attach any penalty to packets being discarded. However, note that if the per-user throughputs are all equal to $T/N$ (due to symmetry), then the PDR is simply the difference between (per-user) arrival rate and throughput, $\lambda/N - T/N$.

\section{RL Environment and DQN Architecture}\label{sec:rl_env}
We resort to the tool-set of DRL and use DQN \nocite{mnih2015humanlevel} to tackle the problem of RA in MTC networks.
\subsection{Environment}
\begin{figure}
    \centering
    \includegraphics[width=240px, height=110px]{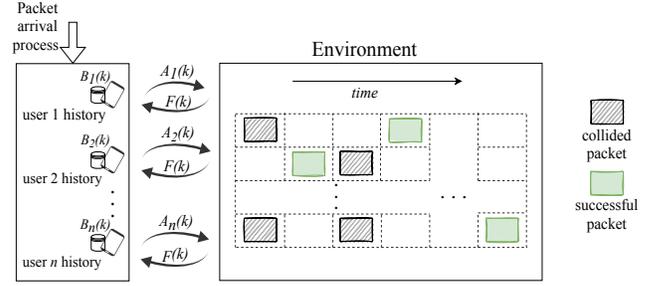}
    \caption{Interaction of agents with the environment. At the beginning each time slot $k$, each agent $n$ first performs an action $A_n(k)$, then receives the feedback signal $F(k)$ at the end of the time slot $k$ from the receiver. Users update their buffers depending on the feedback signal and new packet arrivals
    } \label{fig:MURL_env2}
\end{figure}

The \emph{environment} is the available physical resource in this case, i.e., the channel, as shown in Fig. \ref{fig:MURL_env2}. Every user interacts with the environment by taking an action and receiving a reward signal.
\subsection{State and actions}
By \emph{state} we mean the memory content at user $n$ at time $k$. In this context, we define the state as the local history $S_n(k)$ of each user.\footnote{We will use the terms \textit{history} and \textit{state} interchangeably in the rest of the paper.} The environment is partially observable to each user, i.e., each user $n$ is unaware of the history of the other users. The action $A_n(k)$ of a user $n$ is to transmit $A_n(k)=1$, or wait $A_n(k)=0$.

\subsection{Reward}
Let $R_n(k) \in \mathbb{R}$ be the \textit{immediate} reward that user $n$ obtains at the end of time slot $k$. The reward depends on the user $n$ action $A_n(k)$ and other users' actions $A_{n'}(k)$, $n' \neq n$. The accumulated discounted reward for user $n$ is defined as
\begin{align}
    \mathcal{R}_n(k) = \sum\limits_{k'=0}^{\infty}\gamma^{k'} R_n(k+k'+1),
\end{align}
where $\gamma \in (0,1]$ is a discount factor.

We consider the reward function $R_n(k)$ as successful transmissions. The reward $R_n(k)$ at time slot $k$ is calculated as:
\begin{align} \label{eq:G_reward}
    R_n(k) &= 
    \begin{cases}
		1 & \text{if transmission is successful} \\
		0 & \text{otherwise.}
	\end{cases} 
\end{align}
The reward  is global, i.e., all users receive the same reward, which indicates that the agents are fully cooperative.

Note that in this work we are not interested in optimizing the AoP, nor do we incorporate the AoP into reward function, which could be considered to optimize AoP. We merely use average AoP to assess the fairness achieved by different policies.

\subsection{Deep Q-Network (DQN)}

In many RL algorithms, the basic idea is to estimate the action-value function or Q-function $Q(a,s)$ by using Bellman equation and iteratively updating the Q-values at each time step in the following way:

\begin{align}
    Q_n(a, s) & \xleftarrow{} 
    Q_n(a, s)  
 + \alpha \Big[r_n
     + \gamma \max_{a'} Q_n\big(a', s'\big) 
 - Q_n(a, s) \Big], \nonumber
\end{align}
where $Q(s,a)$ is the old value and and $r_n + \gamma \max_{a'} Q_n\big(a', s'\big)$ is the learned value obtained by getting the reward $r$ after taking the action $a$ at state $s$, moving to the next state $s'$ and then taking the action $a'$ that maximizes $Q_n(a',s')$. $0<\alpha \leq 1$ is the learning rate. Commonly, a function approximate is used to estimate the $Q(a,s)$. In the DQN \cite{mnih2015humanlevel} algorithm, a neural network is used to estimate the $Q(a,s)$. 

\begin{figure}[t]
\centering
    \includegraphics[width=230px, height=130px]{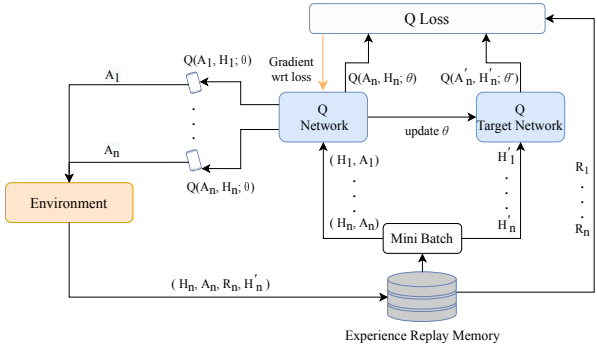}
    \caption{DQN training schematic showing policy network, target network and experience replay.}
    \label{fig:DQN_schematic}
\end{figure}

We use multi-agent DRL and incorporate parameter sharing method \cite{gupta_2017} to perform training in a centralized way for a common policy for all agents, using the experiences of all the agents/users simultaneously. In this way, the $Q_n(a,s)$ does not depend on $n$, which is why we may omit the subscript in notation. We use the \textit{experience replay} technique by storing the experience samples $(s,a,r,s')$ of each user in a memory buffer $\mathcal{D}$, and sampling them uniformly as mini-batches of size $M$ from $\mathcal{D}$ for training. Moreover, we use two neural networks as in \cite{doubleDQN}. The Q-network with parameters $\theta$ that is used to evaluate and update the policy, and the target network with parameters $\theta^-$. The parameters of the Q-network are frequently copied to the target network. This process is also depicted in Fig. \ref{fig:DQN_schematic}. At every time step, the current parameters $\theta$ are updated minimizing the Q-loss function
\begin{align*}
    L(\theta) = \frac{1}{M} \sum_{i=1}^{M} \big(r_i+  \gamma \max_{a'} Q(a',s'_i; \theta^-) - Q(a_i,s_i; \theta)\big)^2. 
\end{align*}
The learned common policy $\pi$ is deployed identically over the set of $N$ users, who take actions without coordination.

At each time slot $k$, each user $n$ obtains the observation $F(k)$, updates its history $S_n(k)$ with it and then feeds $S_n(k)$ to the DQN, whose output are the Q-values for all the available actions. User $n$ follows the policy $\pi$ by drawing an action $A_n(k)$ from the following distribution calculated using the softmax policy \cite{Sutton_2E}
\begin{align}
    \pi(a|s)
    &= \frac{e^{\beta Q(a,s)}} {\sum\nolimits_{\tilde{a} \in \{0,1\}} e^{\beta Q(\tilde{a},s)}},
    \qquad \forall a \in \{0,1\},
\end{align}
where $\beta > 0$ is the temperature parameter which is used to adjust the balance between \textit{exploration} and \textit{exploitation}.

\section{Experiments}\label{sec:experiments}
We perform simulations for $N=10$ users. For DQN training, we use a fully connected feedforward neural network with two hidden layers of 30 and 20 neurons in each.

\subsection{DQN Training Setup}
\paragraph{Transfer Learning} We have experienced that the DQN can be trained well if the total arrival rate $\lambda$ is not too large or too small. In the extreme cases, the DQN may not be able to observe and explore all possible states well enough. Therefore, a transfer learning approach is considered. More specifically, once the DQN is trained for a given value of $\lambda$, the weights are transferred to train for lower or  higher values of $\lambda$. In the simulations, we start training the DQN for $\lambda = 0.20$ and use it to train for higher values of $\lambda$.

\paragraph{Adaptive learning rate}
We consider an adaptive learning rate $\alpha$ for the initial $\lambda$. After each iteration (step), we update it as
$\alpha \xleftarrow{} \max \Bigl( \frac{0.01}{5^{\mathrm{step}}}, 10^{-6} \Bigr)$.
During the training, the learning rate is not reset to train each $\lambda$ value and it is kept at its minimum value, $10^{-6}$. To allow exploration, during the training, we gradually increase the value of $\beta = 1$ to $\beta_{\mathrm{max}}=20$ for the initial value of $\lambda$ and it is kept at $\beta_{\mathrm{max}}$ for training of the next lambda values. The lower values of $\beta$ parameter allow for more exploration, where for larger values of $\beta$, $\pi(a|s)$ tends to the greedy policy. The training of the DQN is performed for $5{,}000$ time slots for each $\lambda$ and the evaluation over $K=30{,}000$ time slots. The learning rate is updated after $2{,}000$ time slots, while the weights of the target DQN network are updated every $1{,}000$ time slots. We set the discount factor $\gamma = 0.95$ for all the experiments. 

\subsection{Simulation Results and Discussion}
\begin{figure}[t]
    \centering
    \includegraphics[width=200px, height=140px]{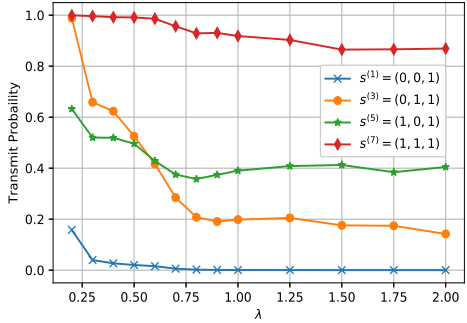}
    \caption{DQN-RA policy transition for each arrival rate $\lambda$. Each legend denotes the state of user $n$ when $B_n(k)=1$.} \label{fig:tr_pr_dqn}
\end{figure}

\begin{figure}[t]
    \centering
    \includegraphics[width=220px, height=140px]{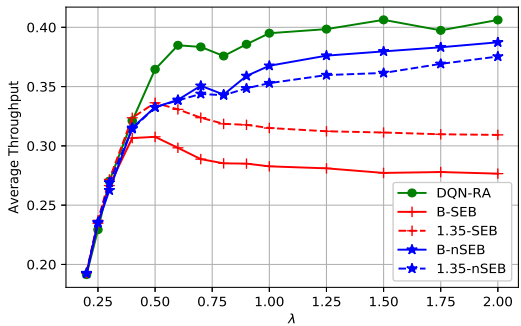}
    \caption{Average Throughput.} \label{fig:th_all_new}
\end{figure}

\begin{figure}[t]
    \centering
    \includegraphics[width=220px, height=140px]{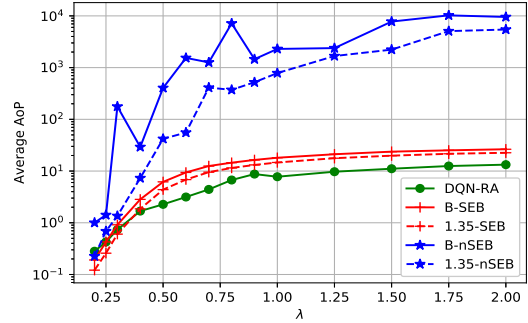}
    \caption{Average AoP in log-scale.} \label{fig:aop_all_new}
\end{figure}

Understanding the behaviour of the transmit probabilities for the different states helps to understand how the DQN learns an efficient policy. The history $S_n(k)$ as defined in \eqref{eq:history_user_n} can take $2^3 = 8$ different values. However, we shall only focus on the states when buffer, $B_n(k) = 1$, that is for each history state $s^{(j)} \in S_n$ for $j = 1, \dotsc, 8$, we define the state $s^{(1)}= (0,0,1)$, $s^{(3)}= (0,1,1)$, $s^{(5)}=(1,0,1)$ and $s^{(7)}=(1,1,1)$. For the other four states $s^{(2)}, s^{(4)}, s^{(6)}, s^{(8)}$, where $B_n(k) = 0$; naturally, the action is $A_n(k)=0$. Fig.~\ref{fig:tr_pr_dqn} depicts how the learned transmit probabilities (policy) vary with the arrival rate $\lambda$. When the state of the user $n$ is $s^{(7)}$, i.e. the last transmission of user $n$ was successful and there is another packet in its buffer, it is evident that at the start, the DQN learns to transmit immediately if a success happens, which is like the immediate-first-transmission (IFT) policy~\cite{Hajek82, kumar1984}. However, for higher arrival rates, it may not be reasonable to transmit as soon as the packet arrives. This is reflected on the transmit probability for $s^{(7)}$, which starts decreasing after $\lambda = 0.6$. Moreover, if user $n$ did not transmit in the last time slot and there was a collision, i.e., state $s^{(1)}$, then the transmit probability almost goes to $0$. The most interesting states are $s^{(3)}$ and $s^{(5)}$. In $s^{(3)}$, when the last transmission was successful but user $n$ did not transmit, as expected, the transmission probability is high for lower arrival rates and it gradually decreases as the arrival rate increases. Moreover, the state $s^{(5)}$ almost remains constant with a transmit probability around $0.4$. These two states provide each user more degrees of freedom to adjust the transmit probability.
The main results of this work are illustrated in Fig.~\ref{fig:th_all_new} and Fig. \ref{fig:aop_all_new}, where it is clear that the proposed DQN-based RA scheme (DQN-RA) learns a policy that outperforms both SEB or nSEB schemes in terms of both throughput and fairness, even though the DQN-RA was not optimized specifically for fairness. One reason is that users have symmetric arrival rates and cooperation among them during centralized training allow them to share the channel in a fair way. The  proposed DQN-RA scheme is compared to the SEB and nSEB for two values of the backoff factor, $\sigma = 1.35$ (1.35-nSEB, 1.35-SEB) and $\sigma = 2$ (B-nSEB, B-SEB). The results show how the throughput--fairness tradeoff varies with $\sigma$. As expected, the nSEB scheme performs better in terms of throughput than the SEB scheme. On the other hand, SEB exhibits a lower average AoP as compared to nSEB. This also shows how different conditions and models can affect the performance of slotted ALOHA.

Furthermore, we have also observed that for the AoP, the nSEB scheme is dependent on the total number of time slots $K$ for which an experiment is evaluated. The AoP of nSEB will increase if $K$ increases because of the \textit{capture effect}. The average AoP becomes higher if a single user occupies the channel for some time. However, the average AoP has moderate values for SEB as well as for the proposed DQN-RA scheme, provided that $K$ is large enough. Please note that we have set $K=100{,}000$ time slots to produce all the results for EB schemes.

\begin{figure*}[t]
    \centering
        \begin{subfigure}{.40\textwidth}
        \includegraphics[width=180px, height=130px]{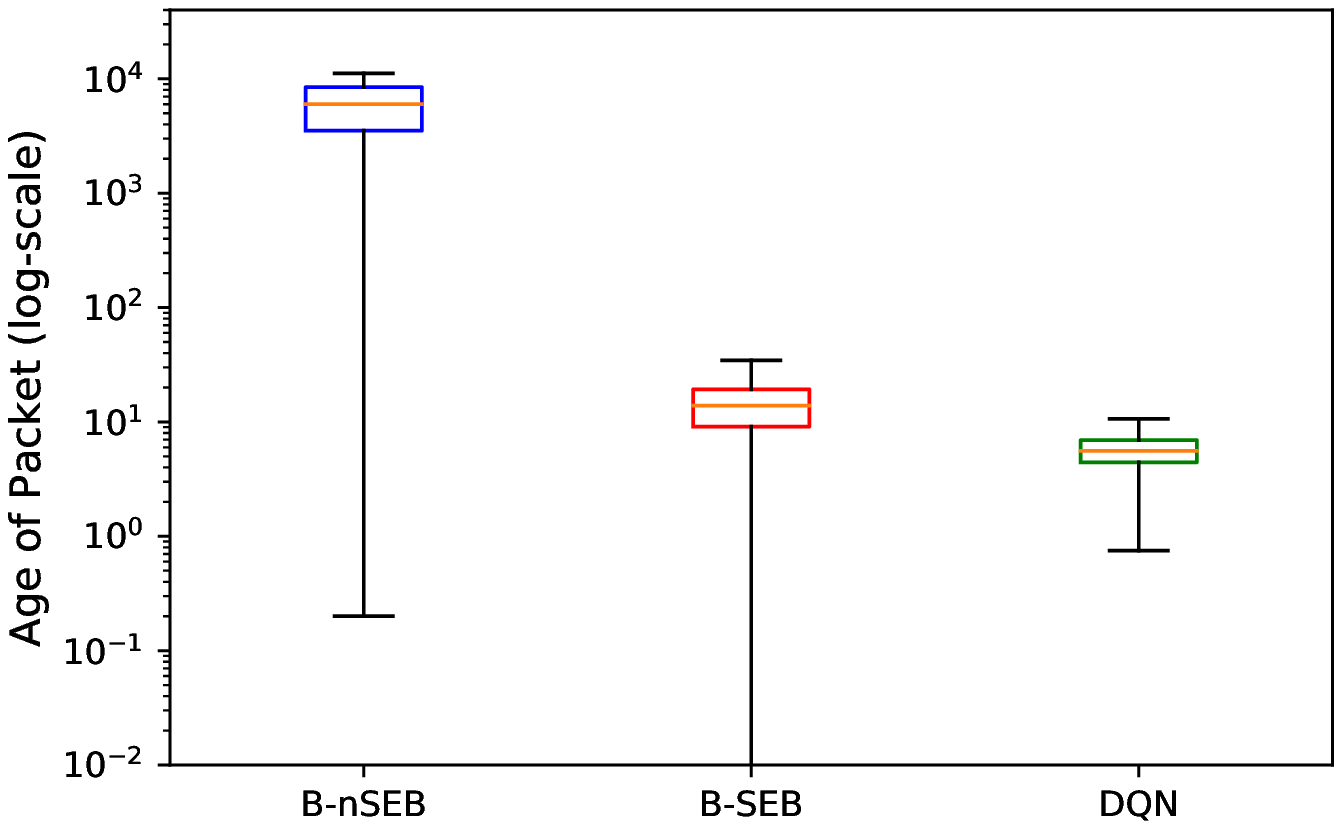}
        \caption{Average AoP of $10$ users}
        \label{fig:aop_box_1_user}
        \end{subfigure}
        \begin{subfigure}{.40\textwidth}
        \includegraphics[width=180px, height=130px]{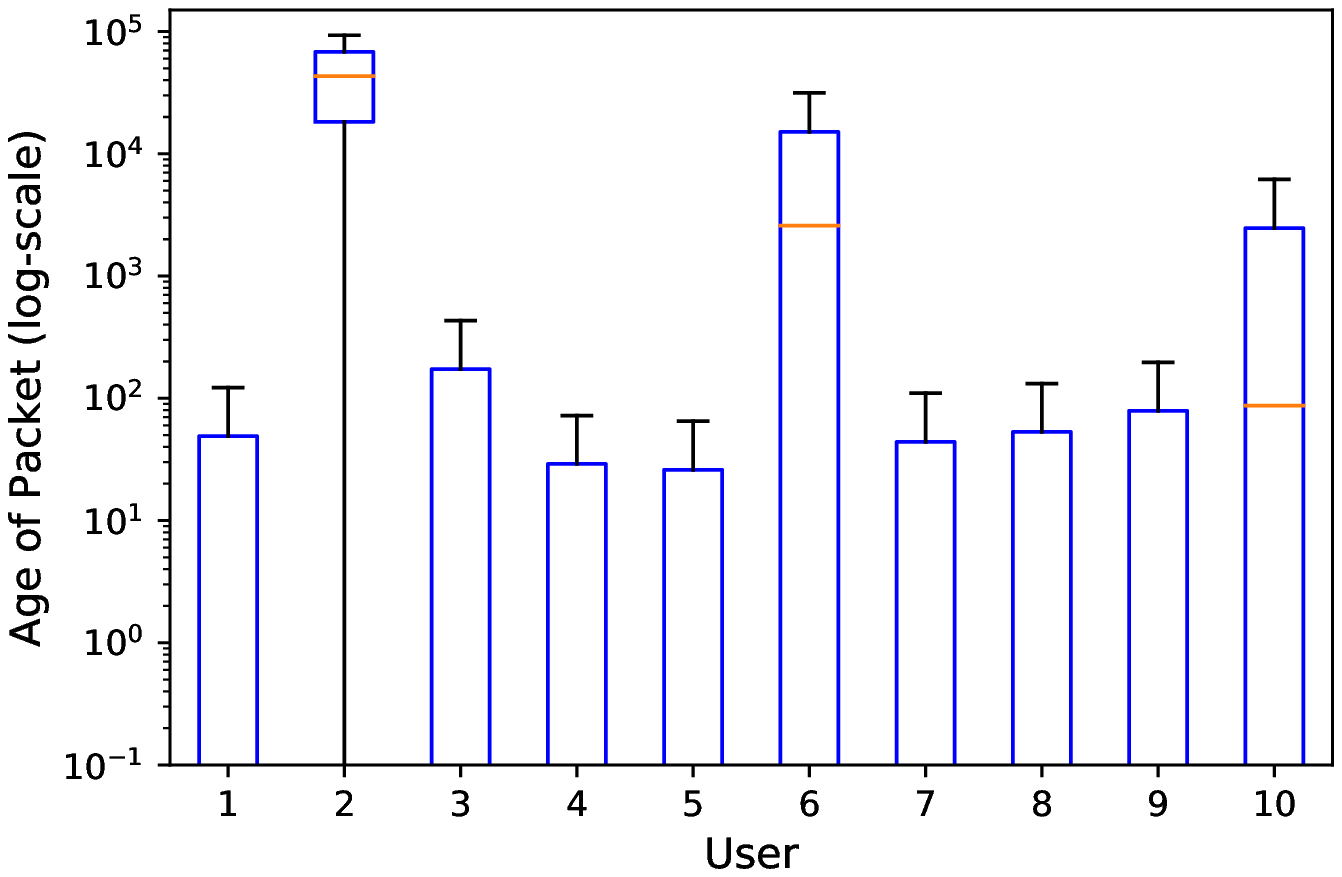}
        \caption{B-nSEB}
        \label{fig:aop_box_beb}
        \end{subfigure}
        \begin{subfigure}{.40\textwidth}
        \includegraphics[width=180px, height=130px]{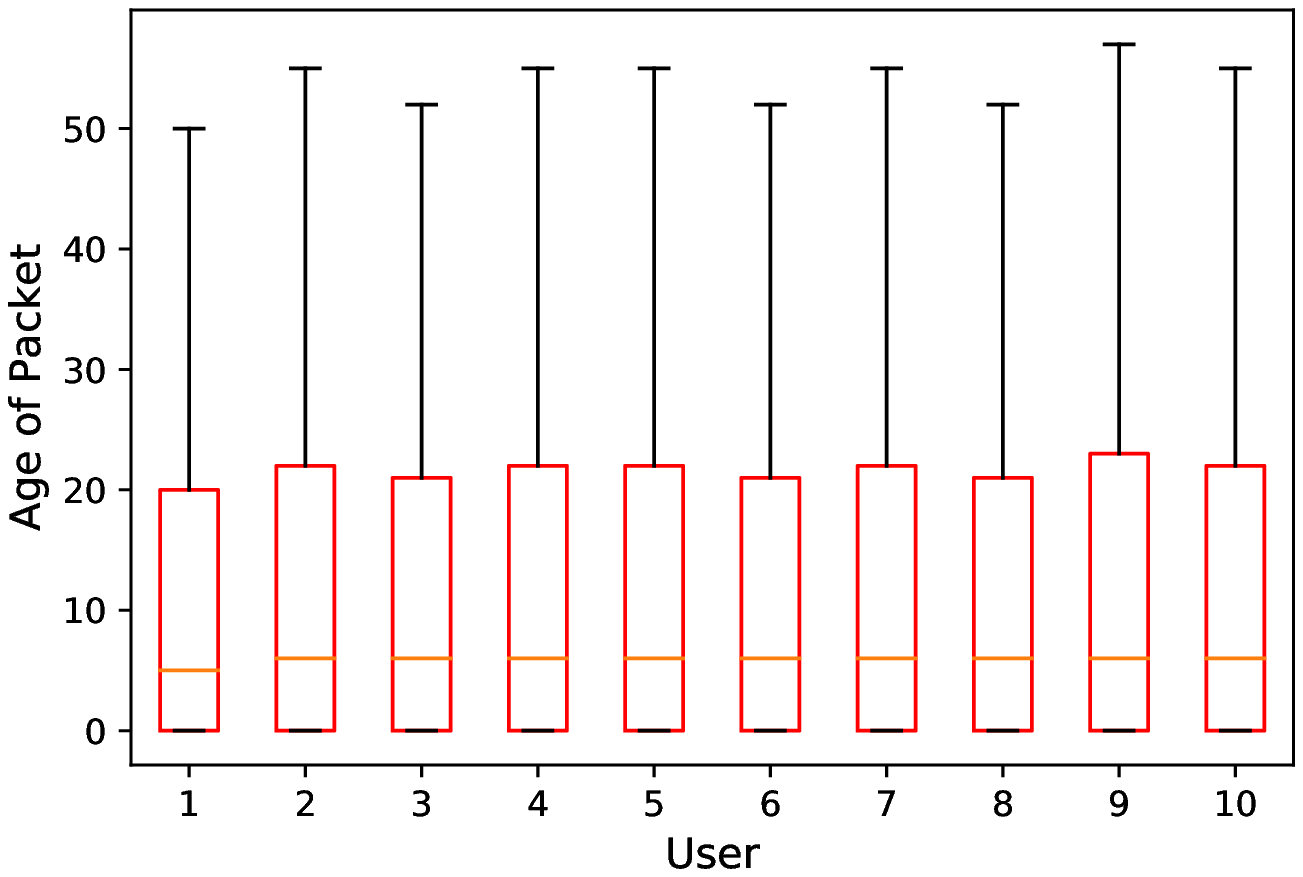}
        \caption{B-SEB}
        \label{fig:aop_box_bseb}
        \end{subfigure}
        \begin{subfigure}{.40\textwidth}
        \includegraphics[width=180px, height=130px]{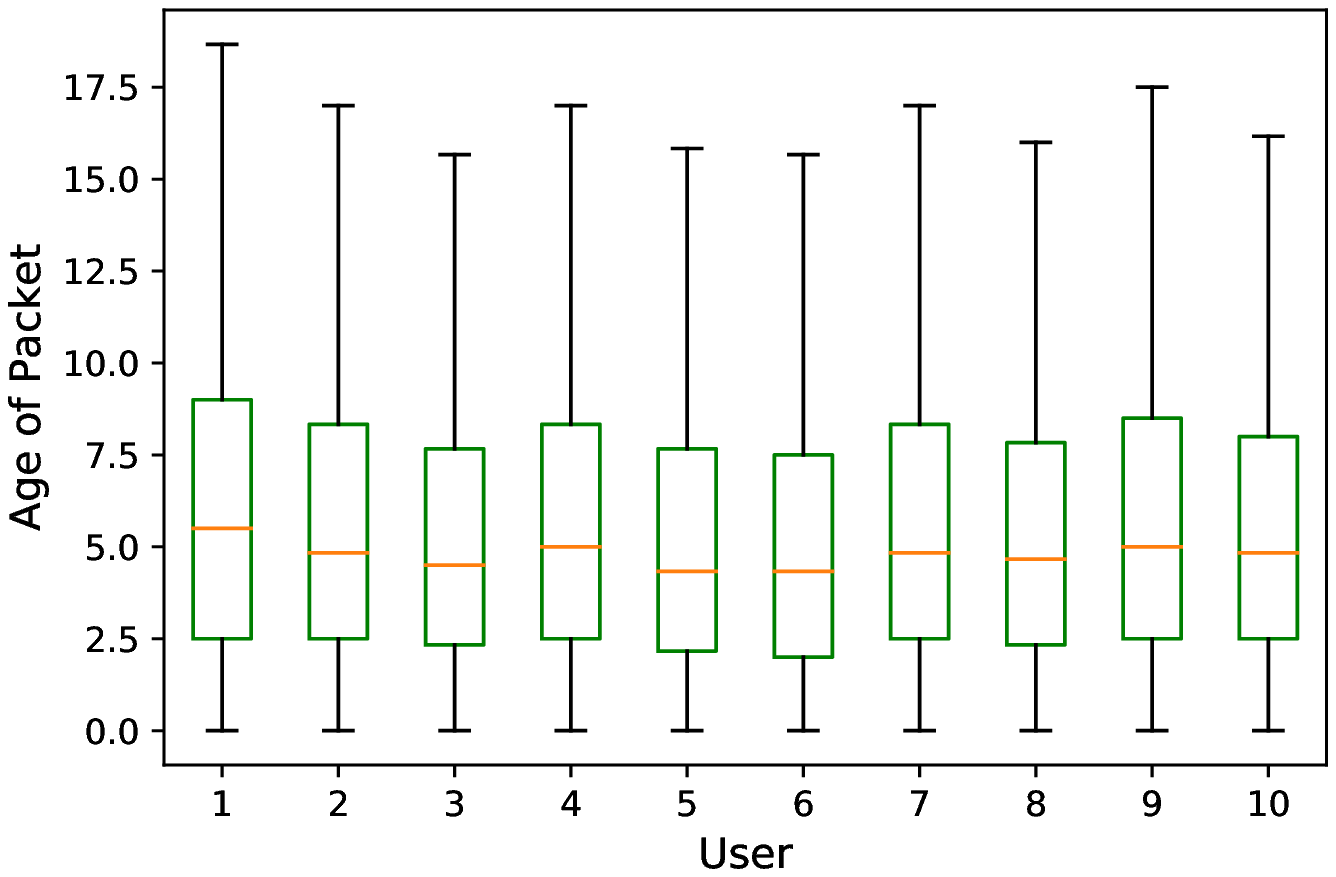}
        \caption{DQN-RA}
        \label{fig:aop_box_dqn}
        \end{subfigure}

    \caption{AoP distribution of all users for the proposed DQN-based schemes and the baseline schemes for $\lambda=0.8$.}
    \label{fig:aop_boxplot}
\end{figure*}

In Fig. \ref{fig:aop_boxplot}, we show standard boxplots for the AoP for the total $N=10$ users over time $K$. The comparison is only shown between the binary EB schemes and the proposed DQN-RA scheme for $\lambda = 0.8$, that has highest AoP for B-nSEB. In Fig. \ref{fig:aop_box_1_user} we show the mean values the 10 users, while in Fig. \ref{fig:aop_box_beb}--\ref{fig:aop_box_dqn}, we can observe the AoP distribution for each user for B-nSEB, B-SEB and DQN-RA, respectively. The unfairness of B-nSEB can be clearly observed from the results where some users, like 4 and 5, are transmitting constantly, while user 2 seldom gets access to the channel; which also leads to significantly larger mean AoP. The results for B-SEB and DQN-RA are more interesting, both B-SEB and DQN-RA have similar medians (i.e., five time slots) for all users. The 25 percentile is 1.0 time slots for the B-SEB and 2.5 for the DQN-RA. The 75 percentile is about 20 time slots for the B-SEB and eight for the DQN-RA. This spread indicates that the DQN-RA is significantly fairer than the B-SEB, as users wait less than eight time slots 75\% of the time for the DQN-RA, while B-SEB users could wait up to 20 time slots. This means that in short burst some B-SEB users will take over the channel and wait only one time slot, while making other users wait for a longer time. This happens frequently enough so every user gets a turn taking over the channel, thus ensuring overall fairness. This overtaking of the channel by one B-SEB user explains why it has the worst throughput of all the methods. The proposed policy does not only achieve the highest throughput, but it is also significantly fairer than all the other methods.

\section{Conclusion and Future Work} \label{sec:conclusion}
In this work, we showed the potential of RL for RA in wireless networks. We proposed a DRL-based transmission policy for RA, DQN-RA for different arrival rates. We proposed to use AoP as a metric to measure fairness of the proposed scheme. Our results showed that the proposed solution learns a policy using previous action and feedback that outperforms standard baseline EB schemes. Moreover, we have also analyzed how DQN-RA scheme's transmission policy changes and adapts to different traffic arrival rates. However, we have not addressed the scalability of this approach to higher number of users, which is our consideration for the next work. For this purpose, we will employ more past actions and feedback signals for learning. 

\section*{Acknowledgment}
This work was supported by the European Union H2020 Research and Innovation Programme through Marie Skłodowska Curie action (MSCA-ITN-ETN 813999 WINDMILL) and the Spanish Ministry of Economy and Competitiveness under Project RTI2018-099722-B-I00 (ARISTIDES).

\bibliographystyle{ieeetr}
\bibliography{bibliography}

\end{document}